\documentclass[aps,prb,reprint,amssymb,amsmath,floatfix]{revtex4-1}

\usepackage[utf8]{inputenc}
\usepackage[T1]{fontenc}

\usepackage{textcomp}
\usepackage{times}
\usepackage{newtxmath}
\usepackage{bm}

\usepackage[english]{babel}
\usepackage{csquotes}

\usepackage{microtype}
\usepackage[colorlinks=true,allcolors=blue]{hyperref}
\usepackage{lipsum}

\usepackage{graphicx}

\usepackage{tikz}
\usepackage{float}

\usepackage{siunitx}
\usepackage{xcolor}
\definecolor{DarkRed}{rgb}{0.65,0,0}
\definecolor{DarkBlue}{rgb}{0,0,0.65}

\renewcommand{\vec}[1]{\bm{#1}}
\providecommand{\drm}{\ensuremath{\mathrm{d}}}
\DeclareMathOperator{\sech}{sech}
\DeclareMathOperator{\real}{Re}
\DeclareMathOperator{\imaginary}{Im}

\providecommand{\eg}{\textit{e.g.}}
\providecommand{\confer}{\textit{cf.}}

\begin{document}

\title{Equations of Motion and Frequency Dependence of Magnon-Induced Domain Wall Motion}
\author{Vetle Risinggård}
\email{vetle.k.risinggard@ntnu.no}
\author{Erlend G. Tveten}
\author{Arne Brataas}
\author{Jacob Linder}
\affiliation{Center for Quantum Spintronics, Department of Physics, NTNU, Norwegian University of Science and Technology, N-7491 Trondheim, Norway}
\date{\today}

\begin{abstract}
Spin waves can induce domain wall motion in ferromagnets. We derive the equations of motion for a transverse domain wall driven by spin waves. Our calculations show that the magnonic spin-transfer torque does not cause rotation-induced Walker breakdown. The amplitude of spin waves that are excited by a localized microwave field depends on the spatial profile of the field and the excitation frequency. By taking this frequency dependence into account, we show that a simple one-dimensional model may reproduce much of the puzzling frequency dependence observed in early numerical studies.
\end{abstract}

\maketitle

\section{Introduction}
Magnon-induced domain wall motion has recently been studied analytically,\cite{Mikhailov1984,Yan2011,*Wang2012,Wang2013a,*Wang2013b,Yan2013} numerically\cite{Hinzke2011,Han2009a,Jamali2010,Seo2011,Wang2012a,Kim2012,Wang2013,Moon2013,Hata2014} and experimentally.\cite{Torrejon2012,Jiang2013} The numerical analyses have uncovered a wide range of domain wall behaviors. The domain wall velocity depends on the frequency of the locally applied magnetic field acting as the spin-wave source in a complicated and non-monotonic way. For some frequencies, it is even possible to reverse the direction of the domain wall motion. This complicated behavior can be explained as a competition between angular and linear momentum transfer: Conservation of angular momentum causes a domain wall to move \textit{towards} the spin-wave source via a magnonic torque.\cite{Mikhailov1984,Yan2011} On the other hand, conservation of linear momentum causes a domain wall to propagate \textit{away} from the spin-wave source.\cite{Wang2013a,*Wang2013b,Yan2013} 

So far, experimental investigations of magnon-induced domain wall motion have mainly focused on dynamics induced by thermal magnons.\cite{Torrejon2012,Jiang2013} However, a domain wall in a temperature gradient can experience additional torques besides the purely magnonic ones, for instance the exchange stiffness can vary with temperature.\cite{Hinzke2011,Kovalev2012a,Kovalev2014a,Wang2014,Schlickeiser2014,Yan2015,Moretti2017}  Characteristically, these torques can induce a Walker breakdown, upon which the domain wall is deformed as it moves.\cite{Schryer1974} 

In ferromagnets, previous studies of domain wall dynamics due to magnonic torques have considered the response of a \textit{static} wall to \textit{first-order} spin-wave excitations. In such a scheme, global conservation laws determine the resulting domain wall velocity.\cite{Mikhailov1984,Yan2011,Wang2013a,Yan2013} However, understanding \textit{dynamic} phenomena such as a Walker breakdown requires knowledge about the dynamics of the collective coordinates that represent the domain wall.\cite{Tretiakov2008} Spin-wave-induced domain wall motion is a result of the back-action of the spin waves on the magnetic texture.\cite{Mikhailov1984} Consequently, the soft modes of the domain wall are quadratic in the spin-wave amplitude. Deriving the equations of motion of the collective coordinates, therefore, requires an expansion to \textit{second order} in the spin-wave amplitudes. This principle is the basis for understanding how spin waves induce domain wall motion in antiferromagnets.\cite{Tveten2014} 

In this article, we apply the same method to ferromagnets. We derive the collective coordinate equations of a spin-wave-driven domain wall from the ferromagnetic Landau--Lifshitz--Gilbert equation. Our approach enables the inclusion of dissipative torques into the dynamic equations of the domain wall position $X$ and the tilt angle $\phi$. 

In the perturbative regime, absence of Walker breakdown amounts to requiring that the domain wall tilt is stationary, $\dot\phi=0$. We show that in practice, the domain wall rotation is always negligibly small in, \eg, YIG. Thus, Walker breakdown is absent in domain wall motion driven purely by magnonic spin transfer.

In our simulations, a localized magnetic field excites the spin-waves that in turn drive the domain wall motion. Understanding the domain wall motion then requires knowing both the frequency-dependent magnonic torques as well as the generated spin-wave amplitude. The spatial profile of the microwave source determines how the spin-wave amplitude depends on the driving frequency.\cite{Kalinikos1981,Gruszecki2016,Korner2017,Fazlali2016} We derive a special case of Kalinikos' general formula\cite{Kalinikos1981} and show that the spin-wave amplitude is proportional to the Fourier sine transform of the source profile. We then use the dependence of the spin-wave amplitude on the microwave frequency to find consistent results for how the domain wall velocity depends on the driving frequency in the numerical and analytical calculations. 

Ref.~\onlinecite{Whitehead2017} has recently considered the linear-response spin-wave emission from a stationary domain wall in a uniform microwave field. We consider a different problem---spin-wave-induced domain-wall motion---which is a second-order effect.

\section{Equations of Motion}

\begin{figure}
  \hspace{-.45cm}\includegraphics[width=.5\textwidth]{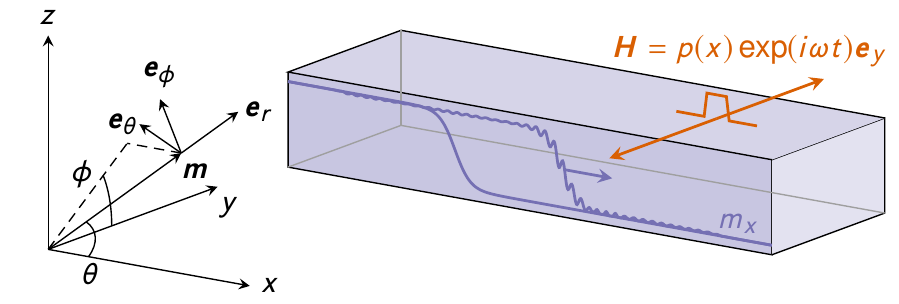}
  \caption{\label{fig:wire} Magnonic spin transfer induces a motion of the domain wall. We consider transverse domain wall motion along the $x$ axis. The spatial profile of the applied magnetic field influences the amplitude of the excited spin waves and subsequently also the resulting domain wall velocity.}
\end{figure}

We consider an effectively one-dimensional ferromagnet, as shown in \autoref{fig:wire}. The Landau--Lifshitz--Gilbert (LLG) equation determines the magnetization dynamics,\cite{Landau1935,*Landau2008,Gilbert2004}
\begin{equation}\label{eq:llg}
\partial_t\vec m=\gamma\vec m\times\vec H+\frac{\alpha}{m}\vec m\times\partial_t\vec m \, ,
\end{equation}
where $\vec m(\vec r,t)$ is the magnetization and $m$ is its magnitude, $\gamma<0$ is the gyromagnetic ratio, $\vec H(\vec r,t)=-\delta F(\vec r,t)/\delta\vec m(\vec r,t)$ is the effective magnetic field and $\alpha>0$ is the Gilbert damping constant.

The free energy $F$ consists of the exchange, the dipole--dipole interaction and the magnetic anisotropy. In a magnetic wire, the dipole--dipole interaction favors a magnetization direction along the long axis. Taking this into account in the simplest approximation, we model the dipole--dipole interaction as an effective easy-axis anisotropy. The free energy is then 
\begin{equation}\label{eq:free}
  F=\!\int\!\!\drm\vec r\!\left(\frac{A(\partial_x\vec m)^2}{m^2}-\frac{Km_x^2}{m^2}\right),
\end{equation}
where $A$ is the exchange stiffness and $K$ is the effective uniaxial anisotropy constant. As shown numerically in Appendix \ref{sect:numerics}, our results are unchanged by an additional hard axis.

The LLG equation~\eqref{eq:llg} conserves the magnitude of the magnetization, $|\vec m|=m$, which makes it convenient to express the magnetization in spherical coordinates (see \autoref{fig:wire}),
\begin{equation}
	\vec m=m(\cos\theta\vec e_x+\cos\phi\sin\theta\vec e_y+\sin\phi\sin\theta\vec e_z)\,.
	\label{eq:mspherical}
\end{equation} 
In terms of the angles $\theta$ and $\phi$ a solution of the static ($\partial_t\vec m=0$) LLG equation \eqref{eq:llg}  is the Néel wall,  $\theta=2\arctan\exp[Q(x-X)/\lambda]$, where $X$ is the domain wall position, $\lambda=\sqrt{A/K}$ is the domain wall width and $Q=\pm1$ is the topological charge.\cite{Shibata2011} In the absence of a hard axis, $\phi$ can take any value.

We calculate the magnon-induced dynamics perturbatively. The small parameter $h$ parametrizes deviations from the equilibrium magnetization in the spherical frame, see \autoref{fig:wire}. To second order in $h$, the magnetization is\cite{Tveten2014}
\begin{align}
\vec m=&\left(m-\frac{h^2}{2m}\left[m_\theta^{(2)}+m_\phi^{(2)}\right]\right)\vec e_r \notag \\
&\qquad\quad+\left(hm_\theta^{(1)}+h^2m_\theta^{(2)}\right)\vec e_\theta+\left(hm_\phi^{(1)}+h^2m_\phi^{(2)}\right)\vec e_\phi \notag \\
=&\left(m-\frac{h^2}{2m}\left[m_\theta^2+m_\phi^2\right]\right)\vec e_r+hm_\theta\vec e_\theta+hm_\phi\vec e_\phi\,. \label{eq:mag}
\end{align}
The second line follows from the the normalization criterion ${\vec m\cdot\vec m}=m^2$ and we have written $m_{\theta/\phi}=m_{\theta/\phi}^{(1)}$ for simplicity. The second order contributions to the transverse components $\vec e_\theta$ and $\vec e_\phi$ have been dropped since it turns out that carrying them through the following calculation does not change Eqs.~\eqref{eq:velocity} and~\eqref{eq:rotation}.

Substituting the expansion~\eqref{eq:mag} into the LLG equation~\eqref{eq:llg} and equating like orders of $h$ gives three equations that determine the magnetization dynamics to zeroth, first, and second order in $h$. Because we are considering the \textit{dynamic} reaction of the magnetization texture, $\theta$ depends on position and both $\theta$ and $\phi$ depend on time. By assuming that the dynamics of the domain wall collective coordinates is quadratic in the spin-wave excitations, we obtain two equations to linear order,
\begin{subequations}\label{eq:mthetaphi}
\begin{equation}
\partial_tm_\theta=\left(-\frac{2\gamma A}{m}\partial_x^2+\frac{2\gamma K}{m}\cos2\theta-\alpha\partial_t\right)m_\phi\,, \label{eq:mtheta}
\end{equation}
and
\begin{equation}
\partial_tm_\phi=\left(+\frac{2\gamma A}{m}\partial_x^2-\frac{2\gamma K}{m}\cos2\theta+\alpha\partial_t\right)m_\theta\,. \label{eq:mphi}
\end{equation}
\end{subequations}
As pointed out in Ref.~\onlinecite{Bayer2005}, the introduction of the auxiliary function $\psi=m_\theta-im_\phi$ simplifies Eqs.\ \eqref{eq:mthetaphi}. Assuming that $\psi(x,t)=\psi(x)\exp(-i\omega t)$ and using $\sin\theta=\sech\xi$, we obtain 
\begin{equation}\label{eq:teller}
q^2\psi=\left(-\partial_\xi^2-2\sech^2\xi\right)\psi \, ,
\end{equation}
where we defined the dimensionless length $\xi=Q(x-X)/\lambda$ and the dimensionless wave number
\begin{equation}
q^2=-m\omega(1+i\alpha)/2\gamma K-1.
\label{eq:q2}
\end{equation}

Eq.~\eqref{eq:teller} is a Schrödinger equation with a reflectionless Pöschl--Teller potential.\cite{Poschl1933} It has two solutions, a bound state $\psi=\rho\sech\xi$ for $q=-i$ (implying $\omega=0$), and a traveling wave\cite{Thiele1973,Lekner2007}
\begin{equation}
\psi(\xi,t)=\rho\left(\frac{\tanh\xi-iq}{1+iq}\right)\exp i(q\xi-\omega t)\,.
\label{eq:psisol}
\end{equation}
The amplitude $\rho$ is arbitrary, and, as is easily checked by back-substitution, the solution~\eqref{eq:psisol} holds for any complex $q$. We need a second equation to determine $m_\theta$ and $m_\phi$. A reasonable condition is that they should both be real, which in turn ensures that the magnetization~\eqref{eq:mag} is real. Thus we write
\begin{subequations}
\begin{align}
m_\theta=&+\real\psi, \\
m_\phi=&-\imaginary\psi.
\end{align}
\end{subequations}

\begin{figure}
\includegraphics[width=.475\textwidth]{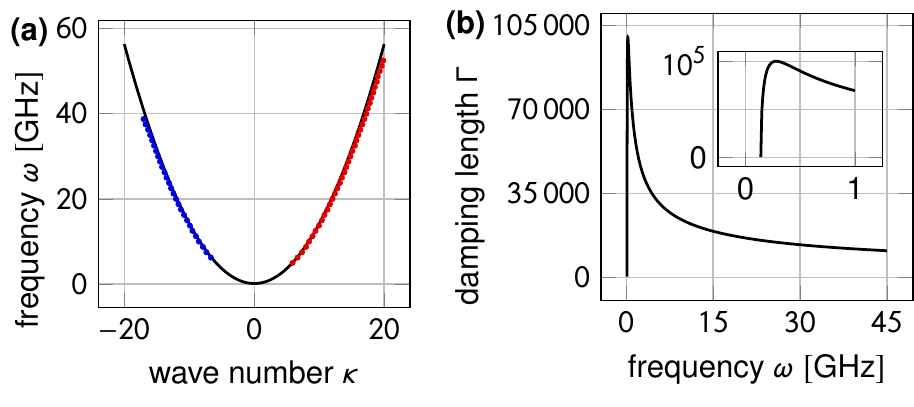}  
\caption{\label{fig:dispersion} (a) Dispersion and (b) damping length of the spin waves to lowest order in $\alpha$. The solid curves are the analytical results of Eqs.~\eqref{eq:dispersion} and~\eqref{eq:dampinglength}. The dotted lines are numerical solutions using a square box (red) and a Gaussian (blue) source. The dimensionless wave number $\kappa$ is given in units of $1/\lambda$ and the dimensionless damping length $\Gamma$ is given in units of $\lambda$. The inset in~(b) shows a zoom-in on $\omega<\SI{1}{\giga\hertz}$. Note that the damping length vanishes as the frequency approaches the gap $\omega_0=-2\gamma K/m=\SI{,14}{\giga\hertz}$ from above. We use material parameters corresponding to YIG: ${\gamma=-\SI{26}{\giga\hertz\per\tesla}}$, $A=\SI{4}{\pico\joule\per\meter}$, $K=\SI{0,4}{\kilo\joule\per\meter\cubed}$, $m=\SI{150}{\kilo\ampere\per\meter}$ and $\alpha=\num{e-5}$.}
\end{figure}

To calculate the real and imaginary part of $\psi$, it is useful to rewrite the wave number $q$ in terms of a real part $\kappa$ and an imaginary part $1/\Gamma$. In Eq.~\eqref{eq:q2}, we insert the real and imaginary part of $q=\kappa+i/\Gamma$ and expand to the lowest non-vanishing order in $\alpha$. We then find the well-known dispersion
\begin{equation}\label{eq:dispersion}
\omega=-\frac{2\gamma K(\kappa^2+1)}{m}
\end{equation}
and damping length\cite{Seo2009}
\begin{equation}\label{eq:dampinglength}
\Gamma=-\frac{4\gamma K\kappa}{m\omega\alpha} \, .
\end{equation}
The dispersion relation and the damping length are both plotted in \autoref{fig:dispersion}. (We use a convention so that $\gamma<0$ and $\alpha>0$.)

The solution~\eqref{eq:psisol} is well known and was used in Ref.~\onlinecite{Yan2011} to derive the domain wall velocity using conservation of angular momentum. It is implicit in the results of Ref.~\onlinecite{Yan2011} that the collective coordinates are quadratic in the spin-wave excitations. Thus, to obtain the main result in this paper, which is the equations of motion for $\dot X$ and $\dot\phi$, we consider the equations that are obtained to second order in $h$ by substituting Eq.~\eqref{eq:mag}~into~Eq.~\eqref{eq:llg},
\begin{subequations}
\begin{align}
  \int\!\!\drm\xi\sech\xi\left(\frac{\dot X}{\lambda}-\alpha\dot\phi\right)\!=&\!\int\!\!\drm\xi\,\frac{4\gamma K\sech\xi}{m^3} \notag\\
  &\times(m_\phi m_\theta\tanh\xi+m_\phi\partial_\xi m_\theta), \label{eq:h2theta} \\
\int\!\!\drm\xi\sech\xi\left(\frac{\alpha\dot X}{\lambda}+\dot\phi\right)\!=&\!\int\!\!\drm\xi\,\frac{4\gamma K\sech\xi}{m^3} \notag\\
  &\times(m_\theta^2\tanh\xi-m_\phi\partial_\xi m_\phi). \label{eq:h2phi}
\end{align}
\label{eq:h2}
\end{subequations}
\hspace{-.25em}Since all terms except $X(t)$ and $\phi(t)$ are known, Eqs.~\eqref{eq:h2} constitute a set of coupled ordinary temporal differential equations that determine the dynamics of the collective coordinates. 

Eqs.~\eqref{eq:h2} contain two different time scales. The fast time scale is set by the period of the spin waves. The slow time scale is associated with the dynamics of the domain wall. Our focus is on the second and slower time scale. Therefore, we substitute the spin-wave components into the right-hand side of Eqs.~\eqref{eq:h2} and average over one spin-wave period, giving 
\begin{widetext}
\begin{subequations}\label{eq:h2t}
\begin{align}
\int\!\!\drm\xi\sech\xi\left(\frac{\dot X}{\lambda}-\alpha\dot\phi\right)\!=&\,\frac{2\gamma A\rho^2\kappa}{m^3\lambda^2}\!\int\!\!\drm\xi\,\exp\left(-\frac{2(\xi+\xi_0)}{\Gamma}\right)\frac{((\kappa^2+1)\Gamma^2+1)\sech\xi+2\Gamma\tanh\xi\sech\xi}{((\kappa^2+1)\Gamma^2+1-2\Gamma)}, \\
\int\!\!\drm\xi\sech\xi\left(\frac{\alpha\dot X}{\lambda}+\dot\phi\right)\!=&\,\frac{2\gamma A\rho^2}{m^3\lambda^2\Gamma}\int\!\!\drm\xi\,\exp\left(-\frac{2(\xi+\xi_0)}{\Gamma}\right)\frac{((\kappa^2-1)\Gamma^2+1)\sech\xi-\Gamma((\kappa^2+1)\Gamma^2-1)\tanh\xi\sech\xi}{((\kappa^2+1)\Gamma^2+1-2\Gamma)}.
\end{align}
\end{subequations}
\end{widetext} 

\noindent Here, the factor $\exp(-2\xi_0/\Gamma)$, where $\xi_0$ is the distance between the domain wall and the spin-wave source, takes into account the damping of the spin-waves. Because the dynamics of the collective coordinates is quadratic in the spin-wave components, their motion decays twice as fast as the spin-waves with increasing distance from the spin-wave source. Carrying out the spatial integrals in Eqs.~\eqref{eq:h2t} then gives the equations of motion. However, direct spatial integration produces beta functions.\cite{Gradshteyn2015} To express the integrals over $\xi$ in terms of elementary functions, we assume that $\xi/\Gamma$ is small and expand the exponential damping factor $\exp(-2\xi/\Gamma)$ on the right-hand side to first order in $\xi/\Gamma$. For moderate $\xi$, this assumption is valid for large $\Gamma$. This is the case for low damping and frequencies comparable to the gap, see \autoref{fig:dispersion}. When $\xi$ is large, the smallness of $\xi/\Gamma$ is unimportant because the error we introduce is suppressed by the hyperbolic secant.

To linear order in $\alpha$, the resulting equations of motion are
\begin{equation}\label{eq:velocity}
\frac{\dot X}{\lambda}=\frac{2\gamma A\kappa\rho^2\exp(-2\xi_0/\Gamma)(\kappa^2+1+2/\Gamma)}{\lambda^2m^3(\kappa^2+1)}
\end{equation}
and
\begin{equation}\label{eq:rotation}
\dot\phi=\frac{2\gamma A\kappa\rho^2\exp(-2\xi_0/\Gamma)(\alpha\kappa^2+\alpha-3\kappa/\Gamma-1/\kappa\Gamma)}{\lambda^2m^3(\kappa^2+1)}\,.
\end{equation}
Eqs.~\eqref{eq:velocity} and \eqref{eq:rotation} are our main analytical results. As expected, in the limit of no damping, $\alpha\to0$ and $\Gamma\to\infty$, we recover the result of Ref.~\onlinecite{Yan2011},
\begin{equation}\label{eq:yanvel}
\frac{\dot X}{\lambda}=\frac{2\gamma A\kappa\rho^2}{\lambda^2m^3}\,. 
\end{equation}
In addition, the equation of motion for $\phi$ gives
\begin{equation}\label{eq:phinodamp}
\dot\phi=0
\end{equation}
in this limit.

\begin{figure}[tbp]
\includegraphics[width=.425\textwidth]{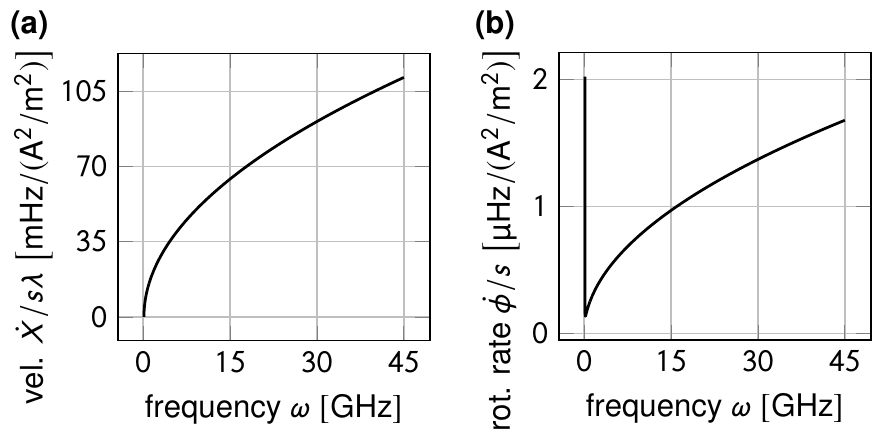}
\caption{\label{fig:collective} The dependence of (a) the domain wall velocity $\dot X/\lambda$ and (b) the rotation rate $\dot\phi$ on the spin-wave frequency. Both plots are normalized to the spin-wave amplitude $s=\rho^2\exp(-2\xi_0/\Gamma)$. The frequency dependence of the wall velocity is monotonic. Because of the low damping, the wall rotation rate is five orders of magnitude smaller than the wall velocity and almost vanishes throughout the interval. We use the same parameters as in \autoref{fig:dispersion}.}
\end{figure}

We have calculated the spin-wave-induced magnetization dynamics perturbatively, \confer\ Eq.~\eqref{eq:mag}. It is then reasonable to assume that the transverse wall will not be transformed into, for instance, a vortex wall.\cite{Beach2008} Thus, absence of Walker breakdown amounts to requiring that the domain wall tilt is stationary, $\dot\phi=0$ (Ref.~\onlinecite{Schryer1974}). Eq.~\eqref{eq:phinodamp} shows that this is always the case for purely magnonic torques. In the presence of a finite damping, $\dot\phi=0$ does not hold identically [\confer\ Eq.~\eqref{eq:rotation}], but when evaluating this expression with material parameters typical of low-damping magnetic garnets (\autoref{fig:collective}) we discover that the rotation rate, although finite, is negligible. Consequently, purely magnonic torques do not induce Walker breakdown in realistic materials.

The dynamics of the domain wall collective coordinates explicitly depends on the frequency of the spin waves through $\kappa$ and $\Gamma$. Assuming the spin-wave amplitude $\rho$ is constant, the domain wall velocity increases monotonically with increasing frequencies, as shown in \autoref{fig:collective}. This is most easily understood in the absence of magnetic damping. Domain wall motion occurs in our model because angular momentum is transferred from the spin waves to the magnetic texture. The group velocity, $v_\text g=\drm\omega/\drm\kappa=4\gamma A\kappa/m\lambda^2$, is a monotonically increasing function of frequency. A higher group velocity implies that more spin waves pass through the domain wall per unit time, so the rate of angular momentum transfer from the spin waves to the domain wall is larger. A higher rate of angular momentum transfer gives a higher domain wall velocity. This is manifest in Eq.~\eqref{eq:yanvel}, which can be written as $\dot X/\lambda=\tfrac{1}{2}v_\text g\rho^2/m^2$.

The monotonic increase in domain wall velocity with increasing frequency contrasts with the non-monotonic dependence on the excitation frequency observed in numerical simulations.\cite{Han2009a,Jamali2010,Seo2011,Wang2012a,Kim2012,Wang2013,Moon2013,Hata2014} However, the dynamics of the collective coordinates also depends strongly on the spin-wave amplitude $\rho$. Since the spin-wave amplitude depends on the frequency of the applied excitation field\cite{Kalinikos1981,Gruszecki2016,Korner2017,Fazlali2016} it is the spatial profile of the applied excitation field that plays the main role in determining the frequency dependence of the domain wall velocity in some of these studies.

\section{Frequency Dependence of the Spin-Wave Amplitude}
We now consider the generation of the spin waves. The microwave source is assumed to be far into the domain. It is then sufficient to only consider the interaction between the source and a homogenous magnetization. Furthermore, we assume that the dominant effect of the damping is an exponential decrease of the spin-wave amplitude as the spin waves move away from the source. This allows us to neglect the damping in the following analysis of the spin-wave generation.

We calculate the disturbance of the homogeneous magnetization caused by the source perturbatively. The small excitation parameter $h$ parametrizes a locally applied magnetic field $\vec H=h\,p(x)\exp(-i\omega t)\vec e_y$. Anticipating a propagating wave solution, we substitute the \textit{ansatz}
\begin{equation}
\vec m=m\vec e_x+hm_y(x,t)\vec e_y+hm_z(x,t)\vec e_z,
\label{eq:mgen}
\end{equation}
which is accurate to first order in $h$, into the LLG equation~\eqref{eq:llg}. This gives two equations to first order in $h$,
\begin{align}
&\partial_tm_y=\left(-\frac{2\gamma A}{m}\partial_x^2+\frac{2\gamma K}{m}\right)m_z\,, \\
&\partial_tm_z=\left(+\frac{2\gamma A}{m}\partial_x^2-\frac{2\gamma K}{m}\right)m_y+\gamma mp(x)\exp(-i\omega t)\,.
\end{align}
Again, introducing the auxiliary variable $\psi=m_y-im_z$ and using  $\psi(x,t)=\psi(x)\exp(-i\omega t)$, we obtain
\begin{equation}\label{eq:gfeq}
-\gamma mp(\xi)=\left(\frac{2\gamma K}{m}\partial_\xi^2-\frac{2\gamma K}{m}-\omega\right)\psi(\xi)\,,
\end{equation}
where we introduced the dimensionless length $\xi=x/\lambda$. To obtain the solution to the differential equation~\eqref{eq:gfeq} for different spatial profiles of the applied magnetic field $p(\xi)$, we solve for the Green function. The Green function $G$ of Eq.\ \eqref{eq:gfeq} is defined by\cite{Boas2006}
\begin{equation}
-\gamma m\delta(\xi)=\left(\frac{2\gamma K}{m}\partial_\xi^2-\frac{2\gamma K}{m}-\omega\right)G(\xi).
\end{equation}
By spatial Fourier transformation, we obtain an algebraic equation that can be solved to give
\begin{equation}
g(\xi')=\frac{\gamma m^2}{2\gamma K(\xi^{\prime2}+1)+m\omega},
\end{equation}
where $\xi'$ is the Fourier conjugate variable of $\xi$. The inverse Fourier transform gives
\begin{equation}
G=-\frac{m^2\sin\kappa\xi}{4\kappa K}\Big(2\Theta(\xi)-1\Big)\,,
\end{equation}
where $\kappa$ is the dimensionless wave number from Eq.~\eqref{eq:dispersion} and $\Theta(\xi)$ is the Heaviside step function. The spin wave $\psi(\xi)$ is then given by the convolution of the Green function and the source profile $p$,
\begin{equation}\label{eq:convolution}
  \psi(\xi)=\!\!\int\limits_{-\infty}^{+\infty}\!\!\drm\xi''\,G(\xi-\xi'')p(\xi'')=2\!\!\int\limits_0^{+\infty}\!\!\drm\xi''\,G(\xi-\xi'')p(\xi''),
\end{equation}
where the last expression, valid only for symmetrical sources $p(-\xi)=p(\xi)$, is proportional to a Fourier sine transform. 

Different source profiles can be obtained by tuning the relative widths of the conducting stripes of a coplanar waveguide.\cite{Gruszecki2016,Korner2017} In particular, we are interested in the square box source $p_1$ and a Gaussian source $p_2$,
\begin{subequations}
\begin{align}
&p_1(\xi)=\frac{H}{2\sigma}\Big(\Theta(\xi-\mu+\sigma)-\Theta(\xi-\mu-\sigma)\Big)\,, \label{eq:square} \\
&p_2(\xi)=\frac{H}{\sqrt{2\pi}\sigma}\exp\left(-\frac{(\xi-\mu)^2}{2\sigma^2}\right), \label{eq:gauss}
\end{align}
\end{subequations}
where $\mu$ is the source position and $\sigma$ is the source half-width. Substituting $p_1$ and $p_2$ into equation \eqref{eq:convolution}, we obtain the spin-wave amplitudes
\begin{subequations}
\begin{align}
&\rho_1=\pm\frac{Hm^2}{4K\kappa}\frac{\sin\kappa\sigma}{\kappa\sigma}, \label{eq:squareamp} \\
&\rho_2=\pm\frac{Hm^2}{4K\kappa}\exp\Big(-\tfrac{1}{2}\kappa^2\sigma^2\Big) \label{eq:gaussamp}
\end{align}
\label{eq:amps}
\end{subequations}
far away from the source ($|\mu|\to\infty$). 

\begin{figure}
\includegraphics[width=.4675\textwidth]{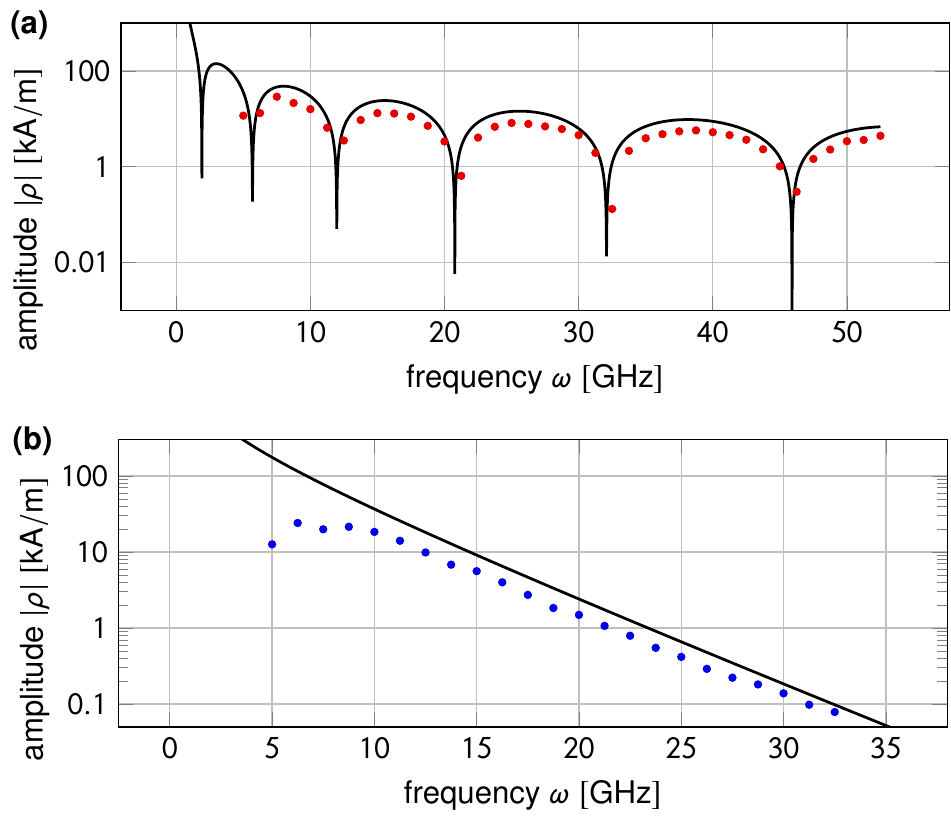}
\caption{\label{fig:amplitude} Spin-wave amplitude as a function of the driving frequency for different spatial profiles of the applied magnetic field. (a)~The amplitude due to the square box source dies off as $|\sin\kappa\sigma/(\kappa^2\sigma)|$, whereas the (b)~the Gaussian source dies off as $\exp(-\kappa^2\sigma^2/2)/\kappa$. We use the same parameters as in \autoref{fig:dispersion} with $H=\SI{0,2}{\tesla}$ and (a)~$\sigma=\num{1,02}\lambda=\SI{102}{\nano\meter}$ and (b)~$\sigma=\lambda/4=\SI{25}{\nano\meter}$. The wave number $\kappa(\omega)$ used to calculate the analytical curves is estimated by doing a least-squares fit to the numerical dispersions in \autoref{fig:dispersion}.}
\end{figure}

Eqs.~\eqref{eq:amps} can be derived as special cases of Kalinikos' general formula.\cite{Kalinikos1981} As illustrated in \autoref{fig:amplitude}, the spin-wave amplitudes depend strongly on the driving frequency and the spatial profile of the applied magnetic field. The square box source \eqref{eq:square} has sharply defined ends. Thus, at every frequency where an integer number of wavelengths fits inside the box, we excite standing waves within the box, and the spin-wave amplitude far away from the box is practically zero. The zeros of the amplitude at \SI{5.63}{\giga\hertz}, \SI{11.90}{\giga\hertz}, \SI{21.25}{\giga\hertz}, \SI{32.50}{\giga\hertz}, and \SI{46.25}{\giga\hertz} correspond to respectively 2, 3, 4, 5, and 6 wavelengths fitting inside the box width of $2\sigma$. The intermittent peaks in the amplitude correspond to a half-integer number of wavelengths fitting inside the box, giving maximum emission of spin waves. On the other hand, the Gaussian source \eqref{eq:gauss} falls off exponentially, and spin waves will leak out of the source region at every frequency. However, the width of the box introduces a length scale in the problem, thus determining the slope of the amplitude in the log-plot found in \autoref{fig:amplitude}(b). Although our analytical calculation readily reproduces the frequency dependence observed in the full numerical solution, it overestimates the amplitude roughly by a factor \num{1,6}. 

\begin{figure}
\includegraphics[width=.4675\textwidth]{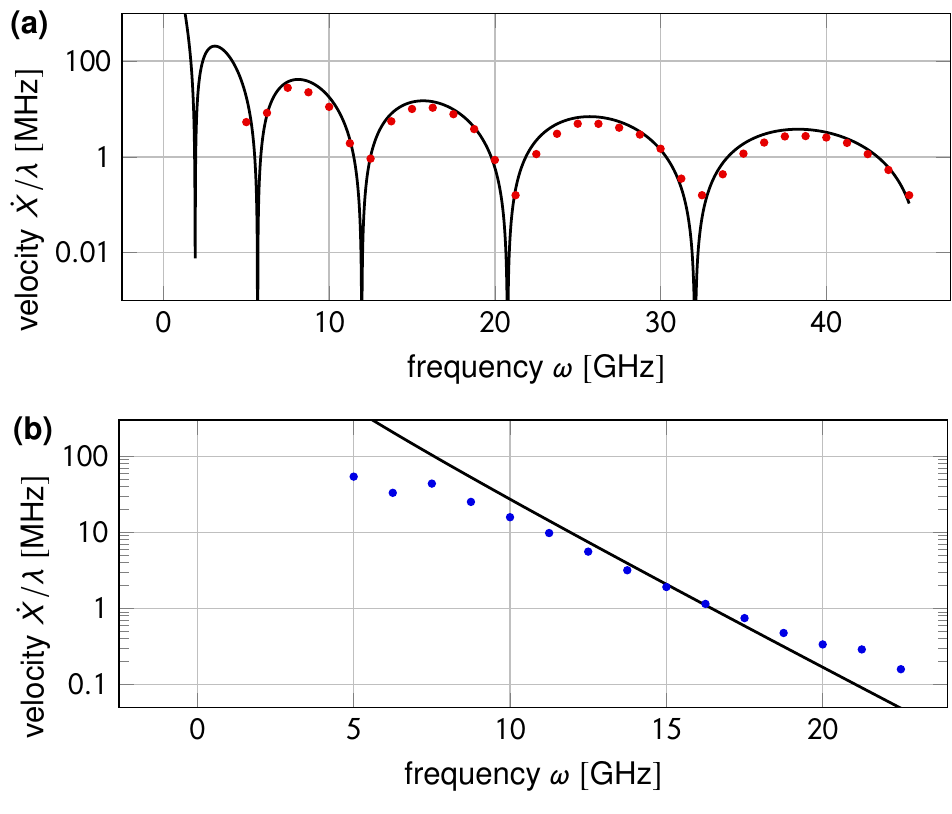}
\caption{\label{fig:velocity} Domain wall velocity as a function of the driving frequency for different spatial profiles of the applied magnetic field. We use the same parameters as in \autoref{fig:dispersion} with $H=\SI{0,2}{\tesla}$ and (a)~$\sigma=\num{1,02}\lambda=\SI{102}{\nano\meter}$ and (b)~$\sigma=\lambda/4=\SI{25}{\nano\meter}$. The analytical curves are calculated by substituting Eqs.~\eqref{eq:amps} into Eq.~\eqref{eq:velocity} and the wave number $\kappa(\omega)$ is estimated by doing a least-squares fit to the numerical dispersions in \autoref{fig:dispersion} and the overestimate of the amplitude has been corrected by a factor $\num{1,6}^2=\num{2,6}$.}
\end{figure}

As shown in \autoref{fig:velocity}, the substitution of the amplitudes \eqref{eq:squareamp} and \eqref{eq:gaussamp} into the equation of motion for $X$, Eq.~\eqref{eq:velocity}, accounts for the frequency dependence of the domain wall velocity in the corresponding one-dimensional numerical model. Appendix~\ref{sect:numerics} describes the numerical calculations.

Our results illustrate that a critical assessment of the impact of the source is vital to extract information about the frequency dependence of magnonic spin-transfer torques from micromagnetic simulations such as those presented in Refs.~\onlinecite{Han2009a,Jamali2010,Seo2011,Wang2012a,Kim2012,Wang2013,Moon2013,Hata2014}. For example, studies that use square box sources need to take into account the well known artifacts\cite{Kalinikos1981} thus introduced in the frequency dependence of the domain wall velocity. The results presented above should serve to illustrate that the frequency dependence introduced by the square box source may account for some of the effects previously attributed to the internal modes of the domain wall.\cite{Han2009a,Seo2011,Wang2012a,Kim2012,Wang2013} However, there are also clear indications that the internal modes of the domain wall affect the magnon-induced domain wall motion when the two-dimensional character of the system is important.\cite{Hata2014} Ref.~\onlinecite{Yan2013a} reports the first steps towards an analytical treatment of magnon--domain wall interaction in two dimensions. 

\begin{figure}
\includegraphics[height=.185\textwidth]{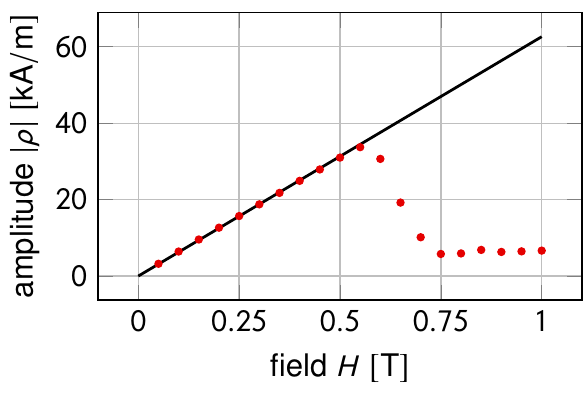}
\caption{\label{fig:field} Spin-wave amplitude as a function of the magnitude of the applied magnetic field $H$ at $\omega=\SI{15}{\giga\hertz}$. We use the square box source ($\sigma=\num{1,02}\lambda=\SI{102}{\nano\meter}$) and the same parameters as in \autoref{fig:dispersion} and the overestimate of the amplitude has been corrected by a factor $\num{1,6}^2=\num{2,6}$.}
\end{figure}

\autoref{fig:field} plots the spin-wave amplitude as a function of the strength of the applied field magnitude $H$. As expected, for small applied fields, there is a linear regime where perturbation theory works well. We observe that a magnetic field of \SI{0,2}{\tesla}, as applied in \autoref{fig:amplitude} and \ref{fig:velocity}, is well within this perturbative regime. For applied magnetic fields above this regime, the amplitude of the mode oscillating at the excitation frequency decreases due to the appearance of higher-frequency modes.\cite{Melkov2013,Demidov2011}

\section{Conclusion}
We have derived the equations of motion for the collective coordinates of a transverse domain wall driven by spin waves. We used this description to demonstrate that magnonic spin transfer does not induce Walker breakdown. For spin waves excited by a localized microwave field the spatial profile of the applied field strongly affects the frequency dependence of the spin-wave amplitude. Taking this frequency dependence into account, we have explained how pure spin transfer may still result in a domain wall velocity with a nonmonotonic dependence on the excitation frequency in a one-dimensional model. In particular, the frequency dependence of the spin-wave amplitude arising from a square box source can account for some of the frequency dependence of the domain wall velocity that have previously been attributed to internal modes of the domain wall. 

\begin{acknowledgments}
We would like to thank Alireza Qaiumzadeh for useful discussions. Funding via the \enquote{Outstanding Academic Fellows} program at NTNU, the COST Action MP-1201, the NV Faculty, the Research Council of Norway Grants No.~216700 and No.~240806, and the Research Council of Norway through its Centres of Excellence funding scheme, Project No.~262633, \enquote{QuSpin} is gratefully acknowledged.
\end{acknowledgments}

\appendix

\section{Numerics}\label{sect:numerics}
We solve the Landau--Lifshitz--Gilbert equation~\eqref{eq:llg} numerically. To this end, we apply a spatio-temporal discretization using a centered implicit scheme in Maple.\cite{Heck2003} The effective field is
\begin{equation}\label{eq:heff}
\vec H=\frac{2A}{m^2}\partial_x^2\vec m+\frac{2}{m^2}(Km_x\vec e_x-K_\perp m_z\vec e_z) \, , 
\end{equation}
as derived from the free energy~\eqref{eq:free} with an additional hard axis anisotropy. The system is a \SI{3}{\micro\meter} grid with grid points spaced \SI{4}{\nano\meter} apart. The initial magnetization profile is a domain wall with positive topological charge ($Q=+1$) and positive chirality ($\phi=0$) centered at the origin. An additional magnetic field $p(x)\exp(-i\omega t)\vec e_y$, centered at $x=\SI{1}{\micro\meter}$, excites spin waves. We insert absorbing boundary conditions at the sample ends to avoid interference phenomena due to spin-wave reflections. In doing so, the Gilbert damping parameter increases to $\alpha=1$ inside \SI{0,3}{\micro\meter} wide regions at both ends of the sample.\cite{Seo2009}

\begin{figure}[b]
\centering
\includegraphics[height=.185\textwidth]{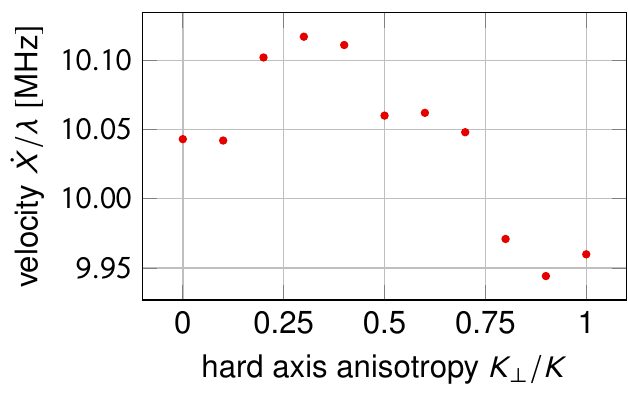}
\caption{\label{fig:perpendicular} Domain wall velocity as a function of hard axis anisotropy for $H=\SI{0,2}{\tesla}$ and $\omega=\SI{15}{\giga\hertz}$. We use the square box source ($\sigma=\num{1,02}\lambda=\SI{102}{\nano\meter}$) and the same parameters as in \autoref{fig:dispersion}.}
\end{figure}

The hard axis anisotropy $K_\perp$ in equation \eqref{eq:heff} is set to zero in the analytical treatment and in the numerical results in the main text. However, we have verified numerically that the results in Figures~\ref{fig:dispersion} and \ref{fig:amplitude}--\ref{fig:field} are essentially unchanged in the presence of a hard axis anisotropy of magnitude $K_\perp=K/2$. As shown in \autoref{fig:perpendicular} for one frequency and applied field magnitude, the additional hard axis anisotropy only leads to slight changes in the domain wall velocity. (This should be expected---in the absence of domain wall rotation the hard axis will not affect the domain wall dynamics.)

%

\end{document}